\documentclass[a4paper]{elsarticle}

\usepackage{lineno,hyperref,amsmath,amssymb,amsthm,verbatim}
\modulolinenumbers[200]

\newtheorem{thm}{Theorem}[section]
\newdefinition{defn}[thm]{Definition}
\newdefinition{example}[thm]{Example}
\newdefinition{lem}[thm]{Lemma}
\newdefinition{rem}[thm]{Remark}
\newdefinition{corollary}[thm]{Corollary}

\journal{ArXiv}

\biboptions{authoryear}

\begin{document}

\begin{frontmatter}

\title{Discrete resonant Rossby/drift wave triads: an explicit parameterisation and a fast direct numerical search algorithm}

%

\author[mymainaddress]{Umar Hayat}
\ead{umar.hayat@qau.edu.pk}

\author[mymainaddress]{Shahid Amanullah}

\author[mysecondaryaddress]{Shane Walsh}

\author[mysecondaryaddress]{Miguel D. Bustamante\corref{mycorrespondingauthor}}

\cortext[mycorrespondingauthor]{Corresponding author}
\ead{miguel.bustamante@ucd.ie}

\address[mymainaddress]{Department of Mathematics, Quaid-i-Azam University, Islamabad, Pakistan}
\address[mysecondaryaddress]{School of Mathematics and Statistics, University College Dublin, Belfield, Dublin 4, Ireland}

\begin{abstract}
We report results on the explicit parameterisation of discrete Rossby-wave resonant triads of the Charney-Hasegawa-Mima equation in the small-scale limit (i.e.~large Rossby deformation radius), following up from our previous solution in terms of elliptic curves \citep{CNSNS}. We find an explicit parameterisation of the discrete resonant wavevectors in terms of two rational variables. We show that these new variables are restricted to a bounded region and find this region explicitly. We argue that this can be used to reduce the complexity of a direct numerical search for discrete triad resonances. Also, we introduce a new direct numerical method to search for discrete resonances. This numerical method has complexity ${\mathcal{O}}(N^3)$, where $N$ is the largest wavenumber in the search. We apply this new method to find all discrete irreducible resonant triads in the wavevector box of size $5000$, in a calculation that took about $10.5$ days on a $16$-core machine. Finally, based on our method of mapping to elliptic curves, we discuss some dynamical implications regarding the spread of quadratic invariants across scales via resonant triad interactions, in the form of sharp bounds on the size of the interacting wavevectors.
\end{abstract}


\end{frontmatter}

\linenumbers

\section{Governing equations}
We shall consider a shallow layer of incompressible fluid, on a rotating planet like the Earth. We may reduce the complications of spherical geometry by using the so-called  $\beta$-plane approximation. Under the consideration of quasigeostropic theory, the dynamics reduce to an equation expressing the conservation of potential vorticity. It is a single partial differential equation for the streamfunction $\psi=\psi(x,y,t)$ (a real scalar field), reading\\
\begin{equation}\label{e:EssIe}
\frac{\partial}{\partial t}(\nabla^2\psi-F\psi)+ \beta \frac{\partial\psi}{\partial x}+\frac{\partial\psi}{\partial x}\frac{\partial \nabla^2\psi}{\partial y}-\frac{\partial\psi}{\partial y}\frac{\partial \nabla^2\psi}{\partial x} =0\,,
\end{equation}
where the value of $\beta$ is constant which shows the speed of rotation of the given system, and $(0 \leq) F=1/L_{R^2}$, where $L_{R}$ is the Rossby deformation radius. The above equation is also known nowadays as Charney-Hasegawa-Mima equation (CHM) because it was independently proposed (several years after the atmospheric context) in the context of plasma physics.

It is easy to find linear wave-like solutions of the CHM equation, of the form $\psi(x,y,t) = \psi_0 \cos \left(\phi_0 + k x+ l y - \omega_{(k,l)} t \right)$, where we introduce the wavevector $(k,l)\in \mathbb{R}^2$ and the frequency $\omega_{(k,l)}$, satisfying the dispersion relation 
$$ \omega_{(k,l)}= -\frac{\beta k}{k^2+l^2+F}\,.$$
This is the well-known Rossby wave formula. In this paper we will consider periodic boundary conditions on the spatial variables $x$ and $y$. Setting $(x,y) \in [0,2\pi)^2$ leads to the following restriction on the wavevectors: $(k,l) \in \mathbb{Z}^2$, which we assume from now on.

When the nonlinear terms in the CHM equation are considered, it is possible to obtain a multi-scale solution in the limit of small amplitudes (i.e. small nonlinearity), of the form:
\begin{equation}
\label{eq:psi_general}
\psi(x,y;t,T) = \Re \sum_{(k,l) \in {\mathcal C}} \Psi_{(k,l)}(T) \exp i \left(k x+ l y - \omega_{(k,l)} t \right)\,,
\end{equation}
 where the set ${\mathcal C}$ is the so-called resonant set, defined as the set of all wavevectors that belong to at least one \emph{resonant triad}, and the functions $\Psi_{(k,l)}(T)$ satisfy a system of nonlinear differential equations in terms of the slow time $T$. 
 
 A resonant triad is, by definition, a triad of wavevectors with integer components: 
 $$(k_1,l_1), (k_2, l_2), (k_3,l_3) \in \mathbb{Z}^2$$ 
 that  solve the so-called triad resonance equations:
\begin{equation}
\label{eq:resonance}
k_1 + k_2 = k_3, \quad l_1 + l_2 = l_3, \quad \omega_1 + \omega_2 = \omega_3\,,
\end{equation}
 where $\omega_j \equiv \omega_{(k_j,l_j)} = -\frac{\beta k_j}{k_j^2+l_j^2+F}\,.$ We remark that the resonance equations are to be satisfied by integers, which makes the problem a Diophantine equation. Solutions exist only if \mbox{$(0\leq) F \in \mathbb{Q}$}. If we set $F=0$ (the so-called small-scale limit) then a mapping to an elliptic surface (and a subsequent mapping to quadratic forms) allows us to parameterise the solutions, as shown for the first time by \cite{CNSNS} and further developed by \cite{Kopp}. See also a particular case discussed independently by \cite{KY}.

We take from here on the small-scale limit case $F=0$ and follow the calculations done in \cite{CNSNS}. We can simplify the system of equations (\ref{eq:resonance}) by eliminating one of the wavevectors: $k_2=k_3-k_1$, $l_2=l_3-l_1$, and writing the equation involving frequencies as a polynomial, which leads to one Diophantine equation for 4 unknowns:

\begin{equation}
\label{eq:reso2}
k_3 (k_1^2+l_1^2)^2 - k_1 (k_3^2+l_3^2)^2  =  2  \left(k_3 (k_1^2+l_1^2) - k_1 (k_3^2+l_3^2)\right)(k_3\,k_1 + l_3\,l_1) \,.
\end{equation}

Notice that this equation is invariant under re-scaling of the wavevectors, which implies the solutions to this equation belong to a projective space. An explicit transformation  from the wavevectors to a new set of variables $(X,Y,D) \in \mathbb{Q}^3$ was done in detail in \cite{CNSNS}:
$$\frac{k_1}{k_3}=\frac{X}{D^2+Y^2}, \qquad  \frac{l_1}{k_3}=\frac{X}{Y}\left(1-\frac{D}{D^2+Y^2}\right), \qquad  \frac{l_3}{k_3}=\frac{D-1}{Y}$$
with $k_3 \neq 0$ (which can be assumed in general). This transformation's inverse is
 $$X=\frac{k_{3}}{k_{1}}\times \frac{k_1^2+l_1^2}{k_3^2+l_3^2}, \qquad Y=\frac{k_3}{k_1}\times \frac{l_1k_3-l_3k_1}{k_3^2+l_3^2}, \qquad D=\frac{k_3}{k_1}\times \frac{k_1k_3-l_1l_3}{k_3^2+l_3^2}\,.$$
 
 Under this transformation, the resonance equation (\ref{eq:reso2}) becomes an equation defining an elliptic surface:
\begin{equation}\label{e:EssIe1}
         Y^2=X^3-2D X^2+2D X-D^2 ,
\end{equation}
and this can be simplified further to
\begin{equation}\label{e:EssIe2}
      \left(\frac{Y}{X}\right)^2 +\left(\frac{D}{X} + X -1\right)^2=X^2-X+1 \,.
\end{equation}
 The LHS of the above equation is a sum of two squares. Similarly, the RHS can be written in diagonal form by introducing the transformation 
 $$X=-\frac{m+n}{m-n}.$$ 
 We get
\begin{equation}\label{e:EssIe3}
  \left(\frac{Y(m-n)^2}{m+n}\right)^2 + \left(\frac{D(m-n)^2}{m+n} +2m\right)^2=3m^2+n^2\,.
\end{equation}
In summary, we have moved from the $4$ variables $(k_1,k_3,l_1,l_3)$ satisfying the quintic Diophantine equation (\ref{eq:reso2}) to the $4$ variables $(m,n,Y,D)$ satisfying the quadratic equation (\ref{e:EssIe3}). In \citep{CNSNS}, we used this quadratic equation to construct resonant triads via Fermat's Xmas theorem about sums of squares. In the next section, we avoid using this theorem by  producing an explicit parameterisation of the quadratic equation. 

\section{An explicit new parameterisation for resonant triads}

Consider equation (\ref{e:EssIe3}) and define the obvious quantities
$$p=\frac{Y(m-n)^2}{m+n}, \qquad q=\frac{D(m-n)^2}{m+n} +2m\,.$$
We obtain 
$$p^2+q^2=3m^2+n^2\,.$$
We can always assume that one of the variables is non zero. Take $n\neq0$, so this quadratic equation becomes
 $$(p/n)^{2}+(q/n)^{2}=3(m/n)^{2}+1.$$
We introduce auxiliary parameters $R, A, B$ via the formulas $p/n=1+R, \,\, q/n=A R\,\, \text{  and }\,\, m/n= B R$. By substituting these values in the last equation we get 
$$(1+R)^{2}+(A R)^{2}=3(B R)^{2}+1).$$
After simplification we get 
$$ R \left( 2+R(1+A^{2}-3B^{2}) \right)=0.$$\\
The case $R=0$ gives the trivial solution $(p=n,q=0,m=0)$ corresponding to resonant triad interactions involving zonal modes. We assume thus $R\neq0$ and obtain  
$$R=\frac{-2} {1+A^{2} -3B^{2} }. $$

By plugging this value of $R$ in $p/n=1+R, \,\, q/n=A R \,\, \text{  and } \,\, m/n= B R$, we get $$p/n=\frac{A^{2}-3B^{2}-1}{1+A^{2} -3B^{2}}, \quad q/n=\frac{-2A}{1+A^{2} -3B^{2}}, \quad m/n=\frac{-2B}{1+A^{2} -3B^{2}}. $$
Finally we recover the general case by identifying numerators and denominator, obtaining 
\begin{equation}
\label{eq:param0}
p=A^{2} -3B^{2}-1, \quad  q=-2A, \quad m=-2B \quad \text{ and }\quad n=1+A^{2} -3B^{2}.
\end{equation}

Our task is to write $X$, $Y$ and $D$ in terms of $A$ and $B$. From equations (\ref{eq:param0}) we obtain
$$m+n=-2B+1+A^{2}- 3B^{2}, \qquad m-n=-2B-1-A^{2}+ 3B^{2},$$
so
$$X=\frac{m+n}{m-n}=-\frac{-2B+1+A^{2}- 3B^{2}}{-2B-1-A^{2}+ 3B^{2}}\,.$$
The equation for $q$ reads now
$$\frac{D(m-n)^{2}}{m+n}+2m=-2A$$
so
$$
D=2\frac{(-A+2B)(-2B+1+A^{2}- 3B^{2})}{(-2B-1-A^{2}+ 3B^{2})^2}
$$
and finally we get $Y$ in terms of $A$ and $B$ from
$$
\frac{Y(m-n)^2}{m+n}=A^{2}-3B^{2}-1
$$
so
$$ Y=\frac{(A^{2}-3B^{2}-1)(-2B+1+A^{2}- 3B^{2})}{(-2B-1-A^{2}+ 3B^{2})^2}.
$$
In summary we get the explicit parameterisation
\begin{eqnarray}
\label{eq:XYDparam}
X &=& -\frac{-2B+1+A^{2}- 3B^{2}}{-2B-1-A^{2}+ 3B^{2}}\,,\\
 Y&=&\frac{(A^{2}-3B^{2}-1)(-2B+1+A^{2}- 3B^{2})}{(-2B-1-A^{2}+ 3B^{2})^2}\,,\\
  D&=& 2\frac{(-A+2B)(-2B+1+A^{2}- 3B^{2})}{(-2B-1-A^{2}+ 3B^{2})^2}\,.
\end{eqnarray}

We can find an inverse of these relations, obtaining for $A,B:$
$$A = \frac{2 \left(D^2-D X+X^2-Y^2\right)}{D^2+4 D (Y-X)+(X-Y)^2}\,, \qquad B = -\frac{D^2-X^2+Y^2}{D^2+4 D (Y-X)+(X-Y)^2}\,.$$
Notice that this inverse is not unique. One way to see this is to notice that the resonance condition $D^2 = -2 D X^2+2 D X+X^3-Y^2$ can be used to obtain a different inverse.

Using the above values of $(X,Y,D)$ in terms of $A$ and $B$, we may get the values of $\frac{k_1}{k_3}$,   $\frac{l_3}{k_3}$ and $\frac{l_1}{k_3}$.  Explicitly we find

\begin{eqnarray}
\nonumber
\frac{k_1}{k_3} &=& \frac{\left(A ^2+B  (2-3 B )+1\right)^3}{\left(A ^2-3 B ^2-2 B +1\right) \left(2 \left(11-3 A ^2\right) B ^2+\left(A ^2+1\right)^2-16 A  B +9 B ^4\right)}\,,\\
\nonumber
\frac{l_3}{k_3} &=& \frac{6 \left(A ^2+A -1\right) B ^2-(A +1)^2 \left(A ^2+1\right)+4 A  B -9 B ^4}{\left(A ^2-3 B ^2-1\right) \left(A ^2-3 B ^2-2 B +1\right)}\,,\\
\nonumber
\frac{l_1}{k_3} &=& \frac{\left(A ^2+B  (2-3 B )+1\right)}{\left(A ^2-3 B ^2-1\right) \left(A ^2-3 B ^2-2 B +1\right) \left(2 \left(11-3 A ^2\right) B ^2+\left(A ^2+1\right)^2-16 A  B +9 B ^4\right)} \times \\
\nonumber
 & & \bigg[ A ^6+2 A ^5+A ^4 \left(-9 B ^2-6 B +3\right)-4 A ^3 \left(3 B ^2+2 B -1\right)+3 A ^2 \left(3 B ^2+2 B -1\right)^2 \\
\label{eq:paramFinal}
 && + 2 A  \left(9 B ^4+12 B ^3+14 B ^2-4 B +1\right)-\left(3 B ^2+1\right)^2 \left(3 B ^2+6 B -1\right) \bigg]\,.
 \end{eqnarray}

The inverse of these, as explained above, is not unique. One inverse reads
\begin{eqnarray}
\nonumber 
A &=& -\frac{2 \left( {k_1^4}  {k_3}- {k_1}^3 \left( k_3^2+ l_3^2\right)+ k_1^2  {k_3} \left( {k_3^2}+2  {l_1^2}+ {l_3^2}\right)- {k_1}  {l_1} \left( {k_3^2}+ {l_3^2}\right) ( {l_1}- {l_3})+ {k_3}  {l_1^4}\right)}{ {k_1^4}  {k_3}+ {k_1}^3 \left( {k_3^2}+ {l_3^2}\right)+2  {k_1^2} \left( {k_3^2}  {l_3}+ {k_3}  {l_1^2}+ {l_3}^3\right)- {k_1}  {l_1} (2  {k_3}- {l_1}) \left( {k_3^2}+ {l_3^2}\right)+ {k_3}  {l_1^4}}\,,\\
\label{eq:paramInverse}
B &=& \frac{-\left( {k_1^2}+ {l_1^2}\right) \left( {k_1^2}  {l_1}- {k_1} \left( {l_1^2}+ {l_3^2}\right)+ {l_1}  {l_1^2}\right)}{ {k_1^4}  {l_1}+ {k_1}^3 \left( {l_1^2}+ {l_3^2}\right)+2  {k_1^2} \left( {l_1^2}  {l_3}+ {l_1}  {l_1^2}+ {l_3}^3\right)- {k_1}  {l_1} (2  {l_1}- {l_1}) \left( {l_1^2}+ {l_3^2}\right)+ {l_1}  {l_1^4}}\,.
\end{eqnarray}

Equations (\ref{eq:paramFinal}) and (\ref{eq:paramInverse}) summarise our explicit parameterisation of the resonant triads' wavevector ratios $(k_1/k_3, l_3/k_3, l_1/k_3) \in \mathbb{Q}^3$ in terms of the new parameters $(A, B) \in \mathbb{Q}^2$. In this context it is worth referring the interested reader to the work by \cite{Kopp}, who found another parameterisation independently.

\subsection{Finding a bounded domain in $(A,B)$ plane}

In \cite{CNSNS} we established that a direct search for resonant triads (using e.g.~brute-force methods or more clever methods via parameterisations) could be simplified quite a bit by imposing some ordering convention on the wavevectors:
\begin{itemize}
\item The first idea is that the streamfunction is real, therefore if a wavevector $(k,l)$ appears in the sum in equation (\ref{eq:psi_general}), then its negative $(-k,-l)$ must also appear, and in fact both contribute with exactly the same energy to the system. In other words, the wavevectors $(k,l)$ and $(-k,-l)$ are indistinguishable. Therefore, it is enough to consider a half plane, and we choose the convention $0 < k \in \mathbb{Z}, \,\, l \in \mathbb{Z}$. Notice that the case $k=0$ is important physically because it corresponds to the so-called zonal modes, but in the context of finding solutions to the resonant equation one can discard this case (solutions with $k=0$ are easy to find). 

\item The second idea is that, in a resonant triad, the roles of $(k_1,l_1)$ and $(k_2,l_2)$ are  interchangeable, so one can introduce a convention to avoid counting twice the same physical triad. Such a convention is, for example, $k_1 \leq k_2$, which was used in \cite{CNSNS}. 

\item The third idea is that the governing equation (\ref{e:EssIe}) is invariant under the mirror symmetry $y \to -y$ of the zonal coordinate, which implies that, given a resonant triad, the mirror image obtained by the mapping $l_j \to -l_j\,, \,\,\,j=1,2,3$ is another resonant triad (physically different). Therefore, in the direct search for triads, one can introduce a convention in order to pick only one resonant-triad representative of the mirror symmetry (and get its mirror image in post-processing). Such a convention is, for example, to impose $l_1 < 0$.
\end{itemize}

We have found, using the \emph{Mathematica} command \texttt{Reduce}, that applying the simple ordering convention $\{0<k_1 \leq k_2 < k_3,  \,\, \,\, l_1 < 0\}$ to the wavevector components given by our parameterisation in equation (\ref{eq:paramFinal}) is equivalent to the following conditions for the new $(A, B)$ parameters:
$$\Omega = \left\{(A,B) \in \mathbb{Q}^2 | - 2 \leq A \leq 2, \,\,\,\, \frac{1}{3}-\frac{1}{3} \sqrt{4+3 A ^2} < B < B_{\max}(A)\right\}\,,$$
where we have introduced the piecewise-differentiable continuous function $B_{\max}(A)$, defined by

$$B_{\max}(A) = \left\{\begin{array}{rl}
-\frac{\sqrt{A ^2-1}}{\sqrt{3}}\,,& \mathrm{if} \quad -2 < A \leq A_1 \quad \mathrm{or} \quad A_2 < A < 2\,,\\
&\\
B_0(A) \,,&\mathrm{if} \quad A_1 < A \leq A_2\,,
\end{array}\right.$$
where $A_1, A_2$ are the two real roots of the polynomial $P(A) = 11 A^4 + 20 A^3 - 16 A - 16$, with $A_1 < 0 < A_2$, and $B_0(A)$ is defined for $A \in [A_1,A_2]$ as the negative real root of the polynomial
$Q_A(B) =
27 B ^6
-126 B ^5
-3 \left(9 A ^2+13\right) B ^4
+12 \left(7 A ^2+4 A +1\right) B ^3
 +\left(9 A ^4+22 A ^2+32 A +13\right) B ^2
 -2 \left(7 A ^4+8 A ^3+14 A ^2+8 A +7\right) B 
  -A ^6-3 A ^4 -3 A ^2 -1\,.$

\begin{figure}[h]
\begin{center}
\includegraphics[height=0.6\textwidth]{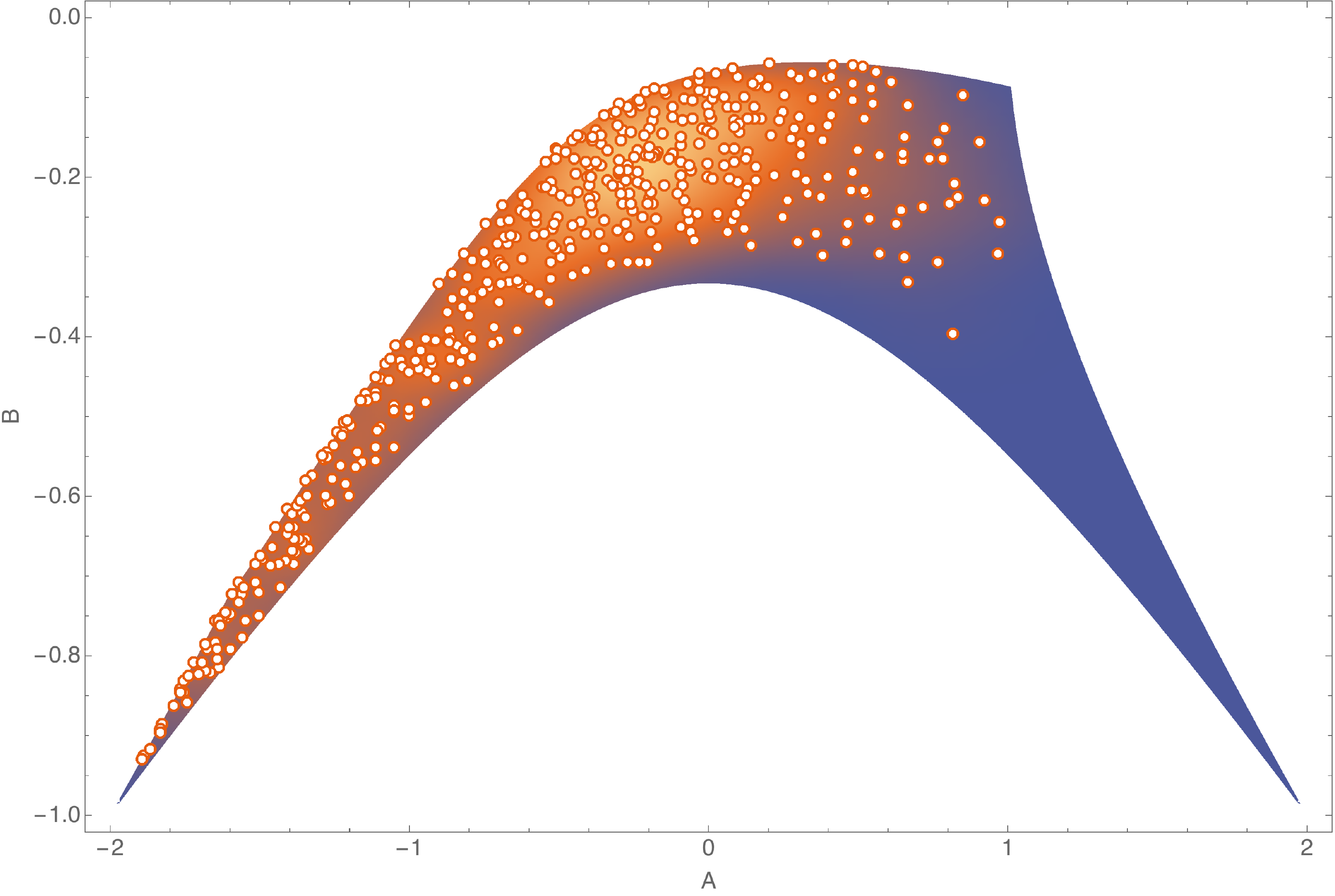}
\includegraphics[height=0.6\textwidth]{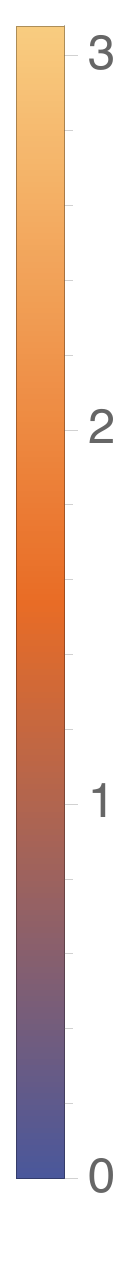}
\caption{Colour online. Open dots denote the values of $(A,B)$ corresponding to all 472 irreducible resonant triads of CHM equation whose wavevectors $(k,l)$ lie within the domain $\{0 < k \leq 5000, \,\,\, -5000\leq l \leq 5000\}$, and where we have used the following convention to avoid repeated counting of triads: $0 <k_1 \leq k_2 < k_3$, and we have selected one representative of the mirror-image symmetry $l\to -l$ by imposing the condition $l_1 < 0$ (thus, there are 944 irreducible triads in reality). The bounded domain in the figure corresponds to these two conditions mapped to the $(A,B)$ plane. The density plot is a smoothed probability density function obtained from the discrete set of points, using a least-squares cross validation method.
\label{fig:ab-plane}}
\end{center}
\end{figure}

Figure \ref{fig:ab-plane} shows a plot of the region $\Omega$ where the parameters $(A,B)$ must lie. The importance of this result is that it establishes that the parameters $(A,B)$ are within a bounded open region, and this can be used to improve direct brute-force searches of resonant triads, with less than ${\mathcal{O}}(N^4)$ complexity, where $N$ is the size of the box in wavevector space. In practice, if we produce grids within the region $\Omega$ having a small enough resolution $\Delta_N$, the resonant triads obtained via equations (\ref{eq:paramFinal}) should be close enough to resonant triads located within a given box of size $N$, and could thus be found by rounding to nearest integers. The question is whether or not we can bound $\Delta_N$ from below, so that the complexity of the calculation can be kept smaller than ${\mathcal{O}}(N^4)$. One would hope to obtain ${\mathcal{O}}(N^3)$ complexity or even lower, but we will leave this discussion to a forthcoming paper.

\section{Improved brute-force numerical search algorithm}
We present the results of our successful attempt to reduce the complexity of a brute-force direct numerical search for triad resonances of the CHM equation. The method uses the resonance equation (\ref{eq:reso2}) to reduce the complexity to ${\mathcal{O}}(N^3)$ where $N$ is the wavevector box size. We implement the code as follows: 
\begin{itemize}
\item Perform 3 nested loops: an outer loop over wavevector component $k_1$ from $1$ to $\lfloor N/2\rfloor$, a middle loop over $k_3$ from $2 k_1$ to $N$ and an inner loop over $l_1$ from $1$ to $N$.
\item Use real numbers whenever possible and parallelise the loops to optimise the code.  
\item Within the loop, solve numerically the resonance equation (\ref{eq:reso2}) for the wavevector component $l_3$ in terms of the components $k_1, k_3, l_1$. There is usually only one real solution. A (quasi-)resonant triad is signalled when the real solution is close enough to an integer, within some tolerance.
\end{itemize}

An example code in \emph{Mathematica} reads as follows:

\begin{verbatim}
Nreso=5001; soluTemp={}; SetSharedVariable[soluTemp]; Clear[l3];
ParallelDo[sq1=1.*(k1^2+l1^2); k3sq1ovk1=k3 sq1/k1;
	uTemp=(l3/.NSolve[k3sq1ovk1 sq1-(k3^2+l3^2)^2+2((k3^2+l3^2)-k3sq1ovk1)(k3 k1+l3 l1)==0,
	l3,Reals]);
	If[NumberQ[Quiet[uTemp[[1]]]], uTemp=Select[uTemp,(Abs[Round[#]-#]<1.*10^(-8))&];
		Do[AppendTo[soluTemp,{{k1,l1},{k3,Round[uTemp[[jj]]]}}],{jj,Length[uTemp]}]],
	{k1,1,Floor[Nreso/2]},{k3,2k1,Nreso},{l1,1,Nreso}];
	soluTemp=Sort[soluTemp];
	Save["file_CHM_triads_compl_N_cube",{Nreso,soluTemp}]
\end{verbatim}

Post-processing work is required on the set of triads obtained:

\begin{itemize}
\item  A selection process is easily applied to ensure that exact resonance is satisfied.
\item Irreducible triads (defined by the condition that the  components $(k_1,k_3,l_1,l_3)$ are coprime) can be easily selected; in view of the projective character of the transformations (\ref{eq:paramFinal}), this is desirable when looking for a unique representative in the $(A,B)$ plane. 
\item As remarked earlier,  this method gives rise to exactly half of the total set of resonant triads in a given wavevector box. The remaining half is obtained in post-processing by applying the mirror symmetry $l \to -l$ to these triads.
\item  Finally, notice that this method controls (by construction) the size of the components $k_1, k_2, k_3, l_1$ to within the box of length $N$, but the size of the $l_2, l_3$ components is not a priori controlled, so an extra selection in post-processing is needed to ensure $|l_2| , |l_3| \leq N$. 
\end{itemize}

As an example, for the box of size $N=5000$ we find all irreducible resonant triads whose wavevectors satisfy $0 < k \leq 5000, \,\,\, -5000\leq l \leq 5000$ and the conventions $0 <k_1 \leq k_2 < k_3$,  $l_1 < 0$. This gives a total of 472 triads (after mirror symmetry is applied, the number of irreducible triads doubles to 944). The calculation took about $10.5$ days on a 16-core machine. These triads are shown explicitly in the Appendix, using the notation $\{k_1,l_1\} \quad \{k_2,l_2\} \quad  \{k_3,l_3\}$ and sorting the triads in terms of smallest $k_1$ and $|l_1|$. We remark the fact that, with such an ordering for the resonant triads, the inequality $l_2>0$ follows necessarily, as can be demonstrated using the explicit parameterisation (\ref{eq:paramFinal}). Physically this implies that resonant interactions always involve positive and negative zonal components, leading to zig-zag patterns.

It is worth mentioning that our search method finds all irreducible resonant triads in a given wavevector box. This shows that there is room for improvement regarding the methods that use explicit parameterisations: \cite{Kopp} found 463 triads (so just 9 triads missing) using his parameterisation and a simple algorithm to scan over the parameter space.

\section{Dynamical issues and spread of quadratic invariants across scales}

We have seen we can assume without loss of generality $k_3 > k_1 > 0$ and $k_3 > k_2 > 0$ but, in contrast to the convention used in the previous section, we will not assume the ordering $k_1\leq k_2$. Rather, in this Section we will order the wavevectors according to their $L^2$ norm: $\|\mathbf{k}_1\| \leq \|\mathbf{k}_2\|\,,$
where $\mathbf{k} \equiv (k,l)$ and $\|\mathbf{k}\| \equiv \sqrt{k^2 + l^2}\,.$
By looking at the frequency resonance condition (\ref{eq:paramFinal}), we see it can be re-written as follows:
$$\|\mathbf{k}_2\|^2 {k_1}(\|\mathbf{k}_3\|^2-\|\mathbf{k}_1\|^2) + \|\mathbf{k}_1\|^2 {k_2}(\|\mathbf{k}_3\|^2-\|\mathbf{k}_2\|^2)  = 0\,,$$
which implies 
$$\|\mathbf{k}_1\| \leq \|\mathbf{k}_3\| \leq \|\mathbf{k}_2\|.$$

At this point it is worth recalling a well-known physical implication stemming from this latter inequality. Regarding the Charney-Hasegawa-Mima partial differential equation, this inequality means that $\mathbf{k}_3$, the wavevector in the resonant triad with the largest meridional component, has a modulus that is in between the moduli of the other two wavevectors in the triad. In physical terms: this wavevector's scale is in between the other two wavevectors' scales. Moreover, a direct inspection of the triad interaction coefficients leads to the conclusion that this mid-scale wavevector is \emph{active} in the sense that the interaction conserves locally the positive-definite quadratic terms of the form
$$|\Psi_{\mathbf{k}_1}|^2 + |\Psi_{\mathbf{k}_3}|^2 , \qquad |\Psi_{\mathbf{k}_2}|^2 + |\Psi_{\mathbf{k}_3}|^2\,,$$ 
where $\Psi_{\mathbf{k}}$ is a Fourier component of the original streamfunction field at wavenumber $\mathbf{k}$: 
$$\Psi_{\mathbf{k}}(t) = \frac{1}{4\pi^2}\int_0^{2\pi}\int_0^{2\pi} \psi(\mathbf{x},t) \exp(- i \mathbf{k}\cdot \mathbf{x})~\text{d}x~\text{d}y\,.$$  

Let us go deeper into the scale relations between the wavevectors in a resonant triad. Motivated by our direct search method to find whether a given wavevector belongs to one (or more) resonant triads, we have studied upper bounds for the largest wavevector modulus, $\|\mathbf{k}_2\|$, and also for the largest wavevector meridional component, $k_3$. We have established the following sharp bounds:
\begin{thm} Consider a resonant triad $\mathbf{k}_1, \mathbf{k}_2, \mathbf{k}_3$ with the ordering convention $0 < k_1 < k_3, \quad 0<k_2<k_3,\quad \|\mathbf{k}_1\| \leq \|\mathbf{k}_3\| \leq \|\mathbf{k}_2\|$. Then
\begin{enumerate}
\item $k_3 \leq  \|\mathbf{k}_3\| \leq \|\mathbf{k}_1\| \frac{r}{U(r^2)}$, where $r \equiv \frac{\|\mathbf{k}_1\|}{k_1} (\geq 1)$ and $U(s)$ is the function determined by the unique real root of the sixth-order polynomial defined by the conditions
$$s^2 - 2\,s(2 \, U^2 - 3 \,U^3 + 2\,U^4) + U^6 = 0,\quad U\geq 1, \quad s\geq1\,.$$
 
\item $\|\mathbf{k}_2\| \leq \|\mathbf{k}_1\| V\left(\frac{r^2}{U(r^2)}\right)$, where $r$ and $U(s)$ are defined in the previous point, and $V(\sigma)$ is the function determined by the unique real root of the fourth-order polynomial defined by the conditions
$$(\sigma-1)(1-2 \, V) - 2\,V^3+ V^4 = 0,\quad V\geq 2, \quad \sigma\geq1\,.$$
\end{enumerate}
\end{thm}

\begin{corollary}
Under the same conditions of the above theorem, the following practical bounds hold:
$$k_3 \leq  \|\mathbf{k}_3\| \leq \|\mathbf{k}_1\| \sqrt{2 \frac{\|\mathbf{k}_1\|}{k_1}}$$
and
$$\|\mathbf{k}_2\| \leq \|\mathbf{k}_1\| \left(1 + \sqrt{1 + 2 \frac{\|\mathbf{k}_1\|}{k_1}}\right)\,.$$
\end{corollary}

These practical bounds provide an understanding of the maximum scale spread across wavevectors in a resonant triad, leading to the scalings
$$(k_3, \|\mathbf{k}_3\|, \|\mathbf{k}_2\|) \propto 
\begin{cases}
\frac{\|\mathbf{k}_1\|^{3/2}}{k_1^{1/2}} &\qquad \text{if} \qquad \frac{\|\mathbf{k}_1\|}{k_1} \gg 1\,,\\
\|\mathbf{k}_1\| &\qquad \text{if} \qquad \frac{\|\mathbf{k}_1\|}{k_1} \gtrapprox 1\,.
\end{cases}
$$

In a forthcoming revision we will provide a full proof of the above Theorem and Corollary, along with a direct numerical assessment of the sharpness of the bounds found.


\bibliographystyle{apalike}
\bibliography{bibliography_CHM}


\section*{Appendix: 472 Irreducible Triads in the box of size 5000. Notation: $\quad \{k_1,l_1\} \quad \{k_2,l_2\} \quad  \{k_3,l_3\}$.} 

$$\left(
\begin{array}{ccc}
 \{1,-8\} & \{15,10\} & \{16,2\} \\
 \{1,-11\} & \{8,34\} & \{9,23\} \\
 \{1,-27\} & \{80,30\} & \{81,3\} \\
 \{1,-47\} & \{135,285\} & \{136,238\} \\
 \{1,-59\} & \{351,183\} & \{352,124\} \\
 \{1,-64\} & \{255,68\} & \{256,4\} \\
 \{1,-83\} & \{159,583\} & \{160,500\} \\
 \{1,-125\} & \{624,130\} & \{625,5\} \\
 \{1,-172\} & \{335,1550\} & \{336,1378\} \\
 \{1,-216\} & \{1295,222\} & \{1296,6\} \\
 \{1,-343\} & \{2400,350\} & \{2401,7\} \\
 \{1,-463\} & \{3104,388\} & \{3105,-75\} \\
 \{1,-512\} & \{4095,520\} & \{4096,8\} \\
 \{2,-49\} & \{190,155\} & \{192,106\} \\
 \{2,-131\} & \{510,925\} & \{512,794\} \\
 \{3,-11\} & \{5,25\} & \{8,14\} \\
 \{3,-11\} & \{13,13\} & \{16,2\} \\
 \{3,-19\} & \{32,44\} & \{35,25\} \\
 \{3,-41\} & \{117,91\} & \{120,50\} \\
 \{3,-89\} & \{325,403\} & \{328,314\} \\
 \{3,-91\} & \{96,404\} & \{99,313\} \\
 \{3,-241\} & \{1440,500\} & \{1443,259\} \\
 \{3,-262\} & \{1024,2108\} & \{1027,1846\} \\
 \{3,-284\} & \{1680,2290\} & \{1683,2006\} \\
 \{3,-389\} & \{3480,2750\} & \{3483,2361\} \\
 \{3,-505\} & \{4536,1542\} & \{4539,1037\} \\
 \{3,-869\} & \{4864,712\} & \{4867,-157\} \\
 \{4,-403\} & \{3200,2450\} & \{3204,2047\} \\
 \{5,-25\} & \{24,62\} & \{29,37\} \\
 \{5,-25\} & \{27,21\} & \{32,-4\} \\
 \{5,-53\} & \{91,169\} & \{96,116\} \\
 \{5,-77\} & \{216,246\} & \{221,169\} \\
 \{5,-325\} & \{48,1106\} & \{53,781\} \\
 \{5,-433\} & \{3016,1334\} & \{3021,901\} \\
 \{6,-163\} & \{640,350\} & \{646,187\} \\
 \{7,-134\} & \{240,550\} & \{247,416\} \\
 \{7,-151\} & \{425,625\} & \{432,474\} \\
 \{7,-181\} & \{153,731\} & \{160,550\} \\
 \{7,-199\} & \{768,824\} & \{775,625\} \\
 \{7,-386\} & \{2288,1586\} & \{2295,1200\} \\
 \{7,-394\} & \{9,176\} & \{16,-218\} \\
 \{7,-719\} & \{2009,451\} & \{2016,-268\} \\
 \{8,-26\} & \{27,51\} & \{35,25\} \\
 \{8,-166\} & \{343,133\} & \{351,-33\} \\
 \{8,-194\} & \{667,319\} & \{675,125\} \\
 \{8,-402\} & \{559,2223\} & \{567,1821\} \\
 \{8,-554\} & \{135,2163\} & \{143,1609\} \\
 \{9,-127\} & \{343,203\} & \{352,76\} \\
 \{9,-253\} & \{535,185\} & \{544,-68\} \\
 \{9,-287\} & \{816,1462\} & \{825,1175\} \\
 \{9,-298\} & \{551,1508\} & \{560,1210\} \\
 \{9,-329\} & \{1271,1681\} & \{1280,1352\} \\
 \{9,-409\} & \{2000,2090\} & \{2009,1681\} \\
 \{9,-461\} & \{416,2314\} & \{425,1853\} \\
 \{9,-497\} & \{2664,1142\} & \{2673,645\} \\
 \{9,-526\} & \{3111,2684\} & \{3120,2158\} \\
 \{9,-682\} & \{2448,578\} & \{2457,-104\} \\
 \{9,-683\} & \{4736,3478\} & \{4745,2795\} \\
 \{9,-733\} & \{3799,1073\} & \{3808,340\} \\
 \end{array}\right)  \left(
 \begin{array}{ccc}
 \{11,-28\} & \{16,58\} & \{27,30\} \\
 \{11,-368\} & \{1573,1594\} & \{1584,1226\} \\
 \{11,-413\} & \{1717,707\} & \{1728,294\} \\
 \{11,-523\} & \{1144,338\} & \{1155,-185\} \\
 \{11,-532\} & \{2277,3134\} & \{2288,2602\} \\
 \{11,-593\} & \{1120,3580\} & \{1131,2987\} \\
 \{11,-611\} & \{2989,3721\} & \{3000,3110\} \\
 \{11,-731\} & \{4320,4452\} & \{4331,3721\} \\
 \{13,-111\} & \{243,231\} & \{256,120\} \\
 \{15,-245\} & \{81,123\} & \{96,-122\} \\
 \{15,-275\} & \{328,1066\} & \{343,791\} \\
 \{15,-295\} & \{17,131\} & \{32,-164\} \\
 \{15,-325\} & \{768,296\} & \{783,-29\} \\
 \{15,-365\} & \{648,246\} & \{663,-119\} \\
 \{15,-400\} & \{1377,1734\} & \{1392,1334\} \\
 \{15,-731\} & \{3625,1537\} & \{3640,806\} \\
 \{16,-38\} & \{35,55\} & \{51,17\} \\
 \{16,-54\} & \{65,78\} & \{81,24\} \\
 \{16,-106\} & \{119,289\} & \{135,183\} \\
 \{16,-158\} & \{39,403\} & \{55,245\} \\
 \{16,-158\} & \{189,498\} & \{205,340\} \\
 \{16,-166\} & \{375,455\} & \{391,289\} \\
 \{16,-250\} & \{609,290\} & \{625,40\} \\
 \{16,-310\} & \{327,1199\} & \{343,889\} \\
 \{16,-310\} & \{1037,986\} & \{1053,676\} \\
 \{16,-562\} & \{71,272\} & \{87,-290\} \\
 \{16,-686\} & \{2385,742\} & \{2401,56\} \\
 \{16,-782\} & \{1859,4732\} & \{1875,3950\} \\
 \{17,-289\} & \{928,754\} & \{945,465\} \\
 \{17,-319\} & \{408,194\} & \{425,-125\} \\
 \{17,-323\} & \{783,327\} & \{800,4\} \\
 \{17,-476\} & \{48,226\} & \{65,-250\} \\
 \{17,-527\} & \{160,1774\} & \{177,1247\} \\
 \{19,-373\} & \{29,899\} & \{48,526\} \\
 \{21,-148\} & \{123,410\} & \{144,262\} \\
 \{21,-727\} & \{2592,948\} & \{2613,221\} \\
 \{23,-401\} & \{96,1132\} & \{119,731\} \\
 \{24,-346\} & \{435,1247\} & \{459,901\} \\
 \{24,-374\} & \{851,1369\} & \{875,995\} \\
 \{24,-514\} & \{1715,1883\} & \{1739,1369\} \\
 \{24,-718\} & \{275,2525\} & \{299,1807\} \\
 \{24,-1207\} & \{2304,722\} & \{2328,-485\} \\
 \{25,-967\} & \{4760,3026\} & \{4785,2059\} \\
 \{26,-247\} & \{64,122\} & \{90,-125\} \\
\{27,-141\} & \{61,343\} & \{88,202\} \\
 \{27,-141\} & \{229,229\} & \{256,88\} \\
 \{27,-174\} & \{53,424\} & \{80,250\} \\
 \{27,-273\} & \{565,791\} & \{592,518\} \\
 \{27,-356\} & \{768,1204\} & \{795,848\} \\
 \{27,-471\} & \{85,1265\} & \{112,794\} \\
 \{27,-736\} & \{208,2366\} & \{235,1630\} \\
 \{27,-924\} & \{4240,2590\} & \{4267,1666\} \\
 \{27,-939\} & \{2173,689\} & \{2200,-250\} \\
 \{27,-1149\} & \{64,538\} & \{91,-611\} \\
 \{27,-1716\} & \{2960,962\} & \{2987,-754\} \\
 \{28,-101\} & \{128,166\} & \{156,65\} \\
 \{28,-289\} & \{512,274\} & \{540,-15\} \\
  \end{array}\right)$$ 
$$ \left(
 \begin{array}{ccc}
 \{29,-418\} & \{403,1456\} & \{432,1038\} \\
 \{29,-522\} & \{1296,2022\} & \{1325,1500\} \\
 \{29,-812\} & \{3315,2950\} & \{3344,2138\} \\
 \{29,-841\} & \{160,412\} & \{189,-429\} \\
 \{29,-841\} & \{1000,470\} & \{1029,-371\} \\
 \{29,-847\} & \{3000,1250\} & \{3029,403\} \\
 \{30,-175\} & \{256,188\} & \{286,13\} \\
 \{31,-467\} & \{729,1737\} & \{760,1270\} \\
 \{32,-596\} & \{625,2275\} & \{657,1679\} \\
 \{32,-844\} & \{513,3189\} & \{545,2345\} \\
 \{32,-906\} & \{1215,4125\} & \{1247,3219\} \\
 \{32,-914\} & \{3055,4225\} & \{3087,3311\} \\
 \{32,-1436\} & \{255,4675\} & \{287,3239\} \\
 \{35,-125\} & \{112,274\} & \{147,149\} \\
 \{35,-305\} & \{429,247\} & \{464,-58\} \\
 \{35,-423\} & \{1053,1209\} & \{1088,786\} \\
 \{37,-796\} & \{1755,3432\} & \{1792,2636\} \\
 \{37,-1036\} & \{2160,750\} & \{2197,-286\} \\
 \{37,-3219\} & \{243,1569\} & \{280,-1650\} \\
 \{39,-143\} & \{81,93\} & \{120,-50\} \\
 \{40,-398\} & \{935,697\} & \{975,299\} \\
 \{40,-610\} & \{303,1919\} & \{343,1309\} \\
 \{42,-661\} & \{2006,1853\} & \{2048,1192\} \\
 \{45,-95\} & \{51,187\} & \{96,92\} \\
 \{45,-1075\} & \{3699,4247\} & \{3744,3172\} \\
 \{46,-507\} & \{1088,1530\} & \{1134,1023\} \\
 \{47,-337\} & \{73,803\} & \{120,466\} \\
 \{47,-968\} & \{3328,1988\} & \{3375,1020\} \\
 \{48,-214\} & \{225,500\} & \{273,286\} \\
 \{48,-3466\} & \{357,1696\} & \{405,-1770\} \\
 \{49,-127\} & \{119,221\} & \{168,94\} \\
 \{49,-1963\} & \{96,908\} & \{145,-1055\} \\
 \{51,-148\} & \{144,278\} & \{195,130\} \\
 \{51,-527\} & \{1280,1000\} & \{1331,473\} \\
 \{51,-1513\} & \{120,710\} & \{171,-803\} \\
 \{51,-1717\} & \{165,4655\} & \{216,2938\} \\
 \{54,-267\} & \{320,250\} & \{374,-17\} \\
 \{54,-699\} & \{1792,1076\} & \{1846,377\} \\
 \{54,-1083\} & \{1802,4505\} & \{1856,3422\} \\
 \{55,-223\} & \{297,419\} & \{352,196\} \\
 \{55,-401\} & \{288,1108\} & \{343,707\} \\
 \{55,-700\} & \{1872,1274\} & \{1927,574\} \\
 \{55,-1076\} & \{2000,802\} & \{2055,-274\} \\
 \{55,-1289\} & \{4913,3587\} & \{4968,2298\} \\
 \{57,-391\} & \{567,1041\} & \{624,650\} \\
 \{57,-704\} & \{1520,790\} & \{1577,86\} \\
 \{58,-667\} & \{1664,1118\} & \{1722,451\} \\
 \{58,-1189\} & \{320,3550\} & \{378,2361\} \\
 \{60,-805\} & \{1604,2807\} & \{1664,2002\} \\
 \{61,-397\} & \{584,1022\} & \{645,625\} \\
 \{61,-487\} & \{920,1250\} & \{981,763\} \\
 \{63,-491\} & \{720,1390\} & \{783,899\} \\
 \{64,-1022\} & \{2970,1725\} & \{3034,703\} \\
 \{65,-403\} & \{544,374\} & \{609,-29\} \\
 \{65,-505\} & \{799,493\} & \{864,-12\} \\
 \{68,-371\} & \{444,925\} & \{512,554\} \\
 \{69,-659\} & \{851,481\} & \{920,-178\} \\
 \{70,-755\} & \{1728,1074\} & \{1798,319\} \\
 \{75,-448\} & \{720,554\} & \{795,106\} \\
 \{75,-968\} & \{592,3034\} & \{667,2066\} \\
 \{76,-1177\} & \{3584,3068\} & \{3660,1891\} \\
 \end{array}\right)
  \left(
 \begin{array}{ccc}
 \{78,-533\} & \{1010,1111\} & \{1088,578\} \\
 \{78,-1859\} & \{306,901\} & \{384,-958\} \\
 \{80,-1102\} & \{175,524\} & \{255,-578\} \\
 \{80,-1442\} & \{481,4366\} & \{561,2924\} \\
 \{81,-192\} & \{175,300\} & \{256,108\} \\
 \{81,-375\} & \{544,510\} & \{625,135\} \\
 \{81,-708\} & \{1200,2050\} & \{1281,1342\} \\
 \{81,-732\} & \{255,374\} & \{336,-358\} \\
 \{81,-951\} & \{1976,3042\} & \{2057,2091\} \\
 \{81,-1029\} & \{2320,1218\} & \{2401,189\} \\
 \{81,-1047\} & \{1824,3652\} & \{1905,2605\} \\
 \{81,-1155\} & \{527,3519\} & \{608,2364\} \\
 \{81,-1536\} & \{4015,1752\} & \{4096,216\} \\
 \{81,-1707\} & \{3360,1300\} & \{3441,-407\} \\
 \{82,-205\} & \{174,377\} & \{256,172\} \\
 \{85,-485\} & \{104,1118\} & \{189,633\} \\
 \{85,-875\} & \{2088,2146\} & \{2173,1271\} \\
 \{87,-1151\} & \{1776,4070\} & \{1863,2919\} \\
 \{87,-1711\} & \{4913,2261\} & \{5000,550\} \\
 \{90,-655\} & \{448,386\} & \{538,-269\} \\
 \{91,-169\} & \{125,235\} & \{216,66\} \\
 \{93,-614\} & \{403,364\} & \{496,-250\} \\
 \{95,-550\} & \{432,1434\} & \{527,884\} \\
 \{95,-605\} & \{256,1528\} & \{351,923\} \\
 \{95,-1150\} & \{2992,3026\} & \{3087,1876\} \\
 \{96,-196\} & \{125,365\} & \{221,169\} \\
 \{96,-428\} & \{629,703\} & \{725,275\} \\
 \{96,-716\} & \{1443,1469\} & \{1539,753\} \\
 \{96,-1588\} & \{2091,1003\} & \{2187,-585\} \\
 \{96,-2012\} & \{4455,1725\} & \{4551,-287\} \\
 \{97,-1067\} & \{416,548\} & \{513,-519\} \\
 \{99,-335\} & \{253,739\} & \{352,404\} \\
 \{99,-743\} & \{1496,1258\} & \{1595,515\} \\
 \{100,-875\} & \{512,484\} & \{612,-391\} \\
 \{101,-404\} & \{256,268\} & \{357,-136\} \\
 \{102,-1739\} & \{4698,2289\} & \{4800,550\} \\
 \{102,-2057\} & \{3328,1378\} & \{3430,-679\} \\
 \{104,-442\} & \{625,775\} & \{729,333\} \\
 \{105,-850\} & \{407,2294\} & \{512,1444\} \\
 \{106,-477\} & \{704,822\} & \{810,345\} \\
 \{109,-1199\} & \{1971,1011\} & \{2080,-188\} \\
 \{111,-259\} & \{145,203\} & \{256,-56\} \\
 \{115,-965\} & \{1029,2863\} & \{1144,1898\} \\
 \{115,-1425\} & \{533,4059\} & \{648,2634\} \\
 \{117,-299\} & \{288,484\} & \{405,185\} \\
 \{117,-871\} & \{315,2225\} & \{432,1354\} \\
 \{117,-1027\} & \{1827,2929\} & \{1944,1902\} \\
 \{119,-289\} & \{256,344\} & \{375,55\} \\
 \{119,-4267\} & \{2401,2177\} & \{2520,-2090\} \\
 \{121,-1457\} & \{3815,2725\} & \{3936,1268\} \\
 \{125,-385\} & \{339,791\} & \{464,406\} \\
 \{125,-390\} & \{243,852\} & \{368,462\} \\
 \{125,-1065\} & \{928,3132\} & \{1053,2067\} \\
 \{128,-406\} & \{246,287\} & \{374,-119\} \\
 \{128,-746\} & \{1300,1375\} & \{1428,629\} \\
 \{128,-842\} & \{522,2233\} & \{650,1391\} \\
 \{128,-874\} & \{1458,1017\} & \{1586,143\} \\
 \{128,-2914\} & \{412,1399\} & \{540,-1515\} \\
 \{129,-1487\} & \{3840,3560\} & \{3969,2073\} \\
 \{130,-545\} & \{638,1189\} & \{768,644\} \\
  \end{array}\right)$$ 
$$ \left(
 \begin{array}{ccc}
\{132,-679\} & \{1020,1445\} & \{1152,766\} \\
 \{132,-2311\} & \{4476,1865\} & \{4608,-446\} \\
 \{135,-300\} & \{208,286\} & \{343,-14\} \\
 \{135,-327\} & \{256,616\} & \{391,289\} \\
 \{135,-625\} & \{936,1118\} & \{1071,493\} \\
 \{135,-1533\} & \{2465,1247\} & \{2600,-286\} \\
 \{135,-1650\} & \{377,806\} & \{512,-844\} \\
 \{135,-1715\} & \{648,874\} & \{783,-841\} \\
 \{135,-1959\} & \{208,902\} & \{343,-1057\} \\
 \{139,-617\} & \{864,1212\} & \{1003,595\} \\
 \{143,-589\} & \{208,322\} & \{351,-267\} \\
 \{146,-1077\} & \{1134,3057\} & \{1280,1980\} \\
 \{147,-1151\} & \{2040,3050\} & \{2187,1899\} \\
 \{149,-512\} & \{315,1150\} & \{464,638\} \\
 \{149,-4023\} & \{256,1848\} & \{405,-2175\} \\
 \{152,-1306\} & \{2071,3815\} & \{2223,2509\} \\
 \{153,-1649\} & \{2520,1310\} & \{2673,-339\} \\
 \{157,-1727\} & \{3843,2257\} & \{4000,530\} \\
 \{159,-583\} & \{736,1028\} & \{895,445\} \\
 \{160,-436\} & \{425,527\} & \{585,91\} \\
 \{160,-748\} & \{1029,1589\} & \{1189,841\} \\
 \{160,-1060\} & \{299,533\} & \{459,-527\} \\
 \{160,-2100\} & \{243,969\} & \{403,-1131\} \\
 \{160,-2675\} & \{3936,1783\} & \{4096,-892\} \\
 \{161,-1877\} & \{4807,4799\} & \{4968,2922\} \\
 \{162,-1239\} & \{1920,1190\} & \{2082,-49\} \\
 \{164,-4141\} & \{1536,2084\} & \{1700,-2057\} \\
 \{165,-775\} & \{1131,1537\} & \{1296,762\} \\
 \{169,-1717\} & \{4160,3530\} & \{4329,1813\} \\
 \{170,-1525\} & \{3286,3763\} & \{3456,2238\} \\
 \{173,-692\} & \{595,544\} & \{768,-148\} \\
 \{176,-602\} & \{625,1225\} & \{801,623\} \\
 \{178,-1691\} & \{1280,980\} & \{1458,-711\} \\
 \{183,-671\} & \{585,1495\} & \{768,824\} \\
 \{183,-3416\} & \{697,1666\} & \{880,-1750\} \\
 \{186,-1007\} & \{1606,1387\} & \{1792,380\} \\
 \{187,-1734\} & \{3888,4134\} & \{4075,2400\} \\
 \{187,-1819\} & \{4272,4094\} & \{4459,2275\} \\
 \{189,-789\} & \{1000,1630\} & \{1189,841\} \\
 \{189,-1002\} & \{1331,1012\} & \{1520,10\} \\
 \{189,-1707\} & \{3731,2717\} & \{3920,1010\} \\
 \{189,-2973\} & \{2275,1625\} & \{2464,-1348\} \\
 \{195,-767\} & \{1029,1141\} & \{1224,374\} \\
 \{205,-910\} & \{1331,1562\} & \{1536,652\} \\
 \{205,-3235\} & \{219,1433\} & \{424,-1802\} \\
 \{205,-3239\} & \{1875,1705\} & \{2080,-1534\} \\
 \{216,-898\} & \{579,2123\} & \{795,1225\} \\
 \{216,-1162\} & \{549,2867\} & \{765,1705\} \\
 \{218,-2071\} & \{4390,2825\} & \{4608,754\} \\
 \{221,-559\} & \{408,1106\} & \{629,547\} \\
 \{221,-697\} & \{259,407\} & \{480,-290\} \\
 \{221,-1469\} & \{1360,4018\} & \{1581,2549\} \\
 \{224,-1124\} & \{901,2809\} & \{1125,1685\} \\
 \{224,-1212\} & \{405,2895\} & \{629,1683\} \\
 \{224,-1562\} & \{2989,2501\} & \{3213,939\} \\
 \{224,-1844\} & \{3645,4653\} & \{3869,2809\} \\
 \{224,-2068\} & \{3451,1943\} & \{3675,-125\} \\
 \{232,-1914\} & \{1955,1275\} & \{2187,-639\} \\
 \{232,-2407\} & \{2840,1619\} & \{3072,-788\} \\
 \{240,-1550\} & \{2847,3139\} & \{3087,1589\} \\
 \{241,-1928\} & \{3087,1876\} & \{3328,-52\} \\
 \end{array}\right)
 \left(
 \begin{array}{ccc}
 \{243,-429\} & \{288,692\} & \{531,263\} \\
 \{245,-1264\} & \{1755,2886\} & \{2000,1622\} \\
 \{247,-1274\} & \{2048,2024\} & \{2295,750\} \\
 \{247,-1924\} & \{3840,2900\} & \{4087,976\} \\
 \{255,-415\} & \{256,488\} & \{511,73\} \\
 \{255,-493\} & \{369,779\} & \{624,286\} \\
 \{256,-462\} & \{324,723\} & \{580,261\} \\
 \{256,-500\} & \{369,820\} & \{625,320\} \\
 \{256,-748\} & \{729,828\} & \{985,80\} \\
 \{256,-956\} & \{986,2057\} & \{1242,1101\} \\
 \{256,-968\} & \{377,2197\} & \{633,1229\} \\
 \{256,-1112\} & \{429,2587\} & \{685,1475\} \\
 \{256,-1144\} & \{1379,2561\} & \{1635,1417\} \\
 \{256,-1364\} & \{2250,2395\} & \{2506,1031\} \\
 \{256,-1372\} & \{2145,1820\} & \{2401,448\} \\
 \{256,-1388\} & \{1890,1425\} & \{2146,37\} \\
 \{256,-1942\} & \{344,917\} & \{600,-1025\} \\
 \{256,-4892\} & \{374,2227\} & \{630,-2665\} \\
 \{257,-509\} & \{351,897\} & \{608,388\} \\
 \{259,-1147\} & \{1616,2222\} & \{1875,1075\} \\
 \{261,-928\} & \{459,578\} & \{720,-350\} \\
 \{264,-1954\} & \{3861,4259\} & \{4125,2305\} \\
 \{267,-890\} & \{1029,1148\} & \{1296,258\} \\
 \{270,-2055\} & \{2048,1384\} & \{2318,-671\} \\
 \{272,-1734\} & \{943,984\} & \{1215,-750\} \\
 \{272,-1994\} & \{3589,4958\} & \{3861,2964\} \\
 \{273,-676\} & \{560,710\} & \{833,34\} \\
 \{275,-925\} & \{768,776\} & \{1043,-149\} \\
 \{279,-953\} & \{992,994\} & \{1271,41\} \\
 \{280,-1006\} & \{287,533\} & \{567,-473\} \\
 \{288,-2948\} & \{4625,2455\} & \{4913,-493\} \\
 \{297,-1641\} & \{1495,4225\} & \{1792,2584\} \\
 \{304,-1658\} & \{2603,3562\} & \{2907,1904\} \\
 \{306,-833\} & \{654,763\} & \{960,-70\} \\
 \{320,-1090\} & \{1246,2047\} & \{1566,957\} \\
 \{320,-1450\} & \{343,3269\} & \{663,1819\} \\
 \{320,-1630\} & \{2058,1589\} & \{2378,-41\} \\
 \{323,-3689\} & \{3952,2314\} & \{4275,-1375\} \\
 \{324,-2163\} & \{4096,4152\} & \{4420,1989\} \\
 \{325,-2431\} & \{3515,2183\} & \{3840,-248\} \\
 \{328,-1510\} & \{2091,1853\} & \{2419,343\} \\
 \{333,-1166\} & \{1443,1708\} & \{1776,542\} \\
 \{333,-1591\} & \{1715,3815\} & \{2048,2224\} \\
 \{333,-3589\} & \{720,1730\} & \{1053,-1859\} \\
 \{336,-1342\} & \{1539,1392\} & \{1875,50\} \\
 \{339,-1243\} & \{480,710\} & \{819,-533\} \\
 \{340,-2363\} & \{3116,2009\} & \{3456,-354\} \\
 \{343,-791\} & \{656,902\} & \{999,111\} \\
 \{343,-1001\} & \{425,625\} & \{768,-376\} \\
 \{343,-1099\} & \{1280,1640\} & \{1623,541\} \\
 \{343,-1351\} & \{432,726\} & \{775,-625\} \\
 \{343,-1652\} & \{1625,4030\} & \{1968,2378\} \\
 \{343,-1834\} & \{377,884\} & \{720,-950\} \\
 \{343,-2387\} & \{4625,4145\} & \{4968,1758\} \\
 \{349,-2792\} & \{816,1402\} & \{1165,-1390\} \\
 \{351,-897\} & \{625,775\} & \{976,-122\} \\
 \{351,-1497\} & \{2080,2804\} & \{2431,1307\} \\
 \{351,-2458\} & \{3825,2516\} & \{4176,58\} \\
 \{352,-1620\} & \{2349,3153\} & \{2701,1533\} \\
 \{352,-1676\} & \{2079,3849\} & \{2431,2173\} \\
 \{361,-1787\} & \{399,4057\} & \{760,2270\} \\
 \end{array}\right)$$ 
$$ \left(
 \begin{array}{ccc}
 \{375,-770\} & \{624,1066\} & \{999,296\} \\
 \{377,-3406\} & \{3207,2138\} & \{3584,-1268\} \\
 \{378,-1149\} & \{902,2419\} & \{1280,1270\} \\
 \{387,-2104\} & \{3501,3890\} & \{3888,1786\} \\
 \{390,-1925\} & \{1088,4726\} & \{1478,2801\} \\
 \{391,-1139\} & \{920,2338\} & \{1311,1199\} \\
 \{395,-1815\} & \{1792,4344\} & \{2187,2529\} \\
 \{395,-1935\} & \{2197,4641\} & \{2592,2706\} \\
 \{405,-1765\} & \{624,4082\} & \{1029,2317\} \\
 \{405,-2454\} & \{4368,4082\} & \{4773,1628\} \\
 \{405,-2865\} & \{1792,1656\} & \{2197,-1209\} \\
 \{407,-1184\} & \{528,754\} & \{935,-430\} \\
 \{408,-2822\} & \{567,1353\} & \{975,-1469\} \\
 \{409,-2045\} & \{2048,1616\} & \{2457,-429\} \\
 \{416,-1378\} & \{613,3065\} & \{1029,1687\} \\
 \{416,-1388\} & \{715,895\} & \{1131,-493\} \\
 \{416,-1638\} & \{475,3675\} & \{891,2037\} \\
 \{423,-2279\} & \{2632,1954\} & \{3055,-325\} \\
 \{424,-2257\} & \{3672,4335\} & \{4096,2078\} \\
 \{429,-1222\} & \{1296,1722\} & \{1725,500\} \\
 \{429,-1261\} & \{696,2726\} & \{1125,1465\} \\
 \{429,-1937\} & \{2448,2074\} & \{2877,137\} \\
 \{432,-1194\} & \{1105,1300\} & \{1537,106\} \\
 \{432,-3666\} & \{1387,1898\} & \{1819,-1768\} \\
 \{440,-790\} & \{561,1037\} & \{1001,247\} \\
 \{456,-1778\} & \{2349,2523\} & \{2805,745\} \\
 \{459,-1229\} & \{901,2491\} & \{1360,1262\} \\
 \{468,-2899\} & \{512,1364\} & \{980,-1535\} \\
 \{477,-1007\} & \{832,1534\} & \{1309,527\} \\
 \{480,-820\} & \{549,1037\} & \{1029,217\} \\
 \{486,-1977\} & \{1664,4602\} & \{2150,2625\} \\
 \{493,-2059\} & \{819,1157\} & \{1312,-902\} \\
 \{495,-3875\} & \{1017,1921\} & \{1512,-1954\} \\
 \{508,-1301\} & \{512,2746\} & \{1020,1445\} \\
 \{511,-3139\} & \{1568,1756\} & \{2079,-1383\} \\
 \{512,-1364\} & \{1372,2149\} & \{1884,785\} \\
 \{512,-2004\} & \{972,1209\} & \{1484,-795\} \\
 \{512,-2236\} & \{3145,4250\} & \{3657,2014\} \\
 \{513,-2019\} & \{760,1130\} & \{1273,-889\} \\
 \{520,-990\} & \{533,861\} & \{1053,-129\} \\
 \{527,-2006\} & \{2560,3700\} & \{3087,1694\} \\
 \{527,-2091\} & \{1984,1782\} & \{2511,-309\} \\
 \{536,-2693\} & \{4096,3658\} & \{4632,965\} \\
 \{544,-1972\} & \{1331,1397\} & \{1875,-575\} \\
 \{544,-2686\} & \{3471,2759\} & \{4015,73\} \\
 \{551,-1972\} & \{2185,2150\} & \{2736,178\} \\
 \{555,-1591\} & \{1632,2788\} & \{2187,1197\} \\
 \{583,-3604\} & \{1232,1874\} & \{1815,-1730\} \\
 \{589,-2047\} & \{1995,4325\} & \{2584,2278\} \\
 \{609,-1160\} & \{768,2036\} & \{1377,876\} \\
 \{609,-1711\} & \{1776,2294\} & \{2385,583\} \\
 \{624,-1066\} & \{671,1769\} & \{1295,703\} \\
 \{625,-1055\} & \{680,1258\} & \{1305,203\} \\
 \{625,-1080\} & \{671,1830\} & \{1296,750\} \\
 \{625,-1715\} & \{1776,2590\} & \{2401,875\} \\
 \{625,-2560\} & \{3471,3560\} & \{4096,1000\} \\
 \{629,-2627\} & \{2091,1927\} & \{2720,-700\} \\
 \end{array}\right)
 \left(
 \begin{array}{ccc}
 \{637,-3341\} & \{1155,1765\} & \{1792,-1576\} \\
 \{638,-3161\} & \{4290,3475\} & \{4928,314\} \\
 \{648,-1446\} & \{715,2925\} & \{1363,1479\} \\
 \{667,-2581\} & \{1725,1775\} & \{2392,-806\} \\
 \{675,-2631\} & \{3485,4633\} & \{4160,2002\} \\
 \{675,-3125\} & \{2997,2479\} & \{3672,-646\} \\
 \{686,-2877\} & \{4050,4875\} & \{4736,1998\} \\
 \{686,-3703\} & \{3840,2900\} & \{4526,-803\} \\
 \{713,-1849\} & \{1495,3575\} & \{2208,1726\} \\
 \{715,-2345\} & \{2552,2726\} & \{3267,381\} \\
 \{720,-1730\} & \{1241,3358\} & \{1961,1628\} \\
 \{721,-2143\} & \{1295,4625\} & \{2016,2482\} \\
 \{725,-1798\} & \{1435,1804\} & \{2160,6\} \\
 \{725,-3700\} & \{1392,1994\} & \{2117,-1706\} \\
 \{729,-2707\} & \{2592,2420\} & \{3321,-287\} \\
 \{729,-2871\} & \{1271,1681\} & \{2000,-1190\} \\
 \{729,-3609\} & \{1280,1928\} & \{2009,-1681\} \\
 \{742,-1537\} & \{1088,1598\} & \{1830,61\} \\
 \{768,-1336\} & \{861,1517\} & \{1629,181\} \\
 \{768,-2264\} & \{2067,2279\} & \{2835,15\} \\
 \{768,-2348\} & \{2562,4087\} & \{3330,1739\} \\
 \{775,-1625\} & \{992,3106\} & \{1767,1481\} \\
 \{803,-2117\} & \{2109,2923\} & \{2912,806\} \\
 \{832,-3302\} & \{918,1717\} & \{1750,-1585\} \\
 \{851,-1369\} & \{864,2076\} & \{1715,707\} \\
 \{864,-2964\} & \{875,1595\} & \{1739,-1369\} \\
 \{870,-2075\} & \{1792,3604\} & \{2662,1529\} \\
 \{901,-2809\} & \{2744,2926\} & \{3645,117\} \\
 \{902,-1681\} & \{1210,2075\} & \{2112,394\} \\
 \{918,-2703\} & \{2410,2651\} & \{3328,-52\} \\
 \{925,-4033\} & \{1235,2119\} & \{2160,-1914\} \\
 \{936,-1963\} & \{1624,2813\} & \{2560,850\} \\
 \{945,-2085\} & \{959,1507\} & \{1904,-578\} \\
 \{952,-4981\} & \{2376,2823\} & \{3328,-2158\} \\
 \{969,-1913\} & \{1496,2638\} & \{2465,725\} \\
 \{1026,-3291\} & \{3838,4949\} & \{4864,1658\} \\
 \{1029,-2291\} & \{1960,2962\} & \{2989,671\} \\
 \{1042,-4689\} & \{1134,2343\} & \{2176,-2346\} \\
 \{1105,-3570\} & \{1296,2058\} & \{2401,-1512\} \\
 \{1107,-3221\} & \{3485,4675\} & \{4592,1454\} \\
 \{1156,-2023\} & \{1404,3003\} & \{2560,980\} \\
 \{1168,-3942\} & \{3773,3759\} & \{4941,-183\} \\
 \{1184,-3580\} & \{1443,2171\} & \{2627,-1409\} \\
 \{1215,-2310\} & \{1233,4384\} & \{2448,2074\} \\
 \{1241,-4307\} & \{1479,2407\} & \{2720,-1900\} \\
 \{1250,-2075\} & \{1374,2977\} & \{2624,902\} \\
 \{1280,-3410\} & \{1372,2149\} & \{2652,-1261\} \\
 \{1280,-3896\} & \{3157,3403\} & \{4437,-493\} \\
 \{1296,-3198\} & \{2697,3451\} & \{3993,253\} \\
 \{1309,-2363\} & \{1616,3838\} & \{2925,1475\} \\
 \{1360,-2482\} & \{1715,2933\} & \{3075,451\} \\
 \{1445,-3500\} & \{3051,4062\} & \{4496,562\} \\
 \{1512,-2634\} & \{1813,3959\} & \{3325,1325\} \\
 \{1631,-3029\} & \{1744,2834\} & \{3375,-195\} \\
 \{1981,-4213\} & \{2264,3398\} & \{4245,-815\} \\
 \{1984,-3394\} & \{2294,4403\} & \{4278,1009\} \\
\end{array}
\right)
$$

~\\

\begin{center}
\textbf{End of Appendix.}
\end{center}

\end{document}